# Is Negative Representation More Engaging? The Influence of News Title Framing of Older Adults on Viewer Behavior

Zhilong Zhao[1] Jiaxin Xia[2]

1. School of Journalism and Communication, South China University of Technology, Guangzhou, China
2. Department of Communication, University of Macau, Macau SAR, China

**Abstract**

Grounded in framing theory, this study examines how the framing of news titles regarding older adults influences user engagement on video-sharing platforms. We analyzed 2,017 video news titles (2016-2021) on Bilibili, a video-sharing platform in China, resulting in nine identified frames with strong inter-coder reliability. Utilizing ANCOVA, we found that negative frames increase views ($F = 3.40$, $p < .001$) and shares ($F = 3.07$, $p < .01$), indicating that negative content related to older adults attracts more attention and encourages further sharing. Conversely, positive frames boost collections ($F = 2.41$, $p < .05$) and rewards ($F = 3.81$, $p < .001$), suggesting that viewers favor and are more likely to provide financial support for positive content. This research highlights the impact of framing in news titles concerning older adults on user engagement and its interplay with ageism. The findings advocate for a more balanced and respectful media portrayal of older adults.

**Introduction**

By 2024, it is estimated that the population aged over 65 will outnumber those under 15 in the WHO European Region (World Health Organization, 2023). Globally, the population aged 65+ is rapidly expanding, projected to reach 16% by 2050 and 24% by 2100 (United Nations Department of Economic and Social Affairs, Population Division, 2022). As the aged population grows, ageism is recognized as a pivotal global issue, affecting millions by perpetuating stereotypes and discrimination against individuals based on their age (World Health Organization, 2021).

Amidst the challenges of an aging population and widespread ageism, there is increasing recognition of the need to transform media narratives (Langmann, 2023). Progressive media and awareness campaigns emphasize portraying older adults in roles that counter stereotypes, highlighting their societal engagement, achievements, and potential (Koskinen et al., 2014; World Health Organization, 2020). However, biased media and online platforms persist, affecting public attitudes and behaviors toward older adults (Ng, 2021; Lytle, Monahan, & Levy, 2023; Ayalon et al., 2021; Zhang & Liu, 2021; Sargu et al., 2023; Reul et al., 2022). Notably, Chang et al. (2020) found that in less-developed countries, 92.7% of ageism-health associations were significant, yet only 8.6% of studies were conducted in these regions.

Cultural norms in China traditionally emphasize respect for older adults, filial piety, and family hierarchy (Yeh et al., 2013). However, rapid modernization has created tension between these traditional values and contemporary ageism (Du, 2013). This dynamic makes media portrayals of older adults in China particularly significant, as they reflect and influence societal attitudes. However, there is a notable research gap in understanding how contemporary media representations of older adults in China may perpetuate ageism, especially among younger generations and on emerging social media platforms.

This study addresses the gap by focusing on a social platform in China. To better understand the relationship between media portrayal of older adults and viewer behavior, we analyzed 2,017 video news titles related to older adults on Bilibili, a Chinese video-sharing platform where users can submit, view, and comment on videos (Walkthechat, September 2, 2021). The study identified nine frames related to the portrayal of older adults in online video news titles. Using ANCOVA, we explored the influence of these frames on user engagement metrics such as views, rewards, collections, and shares, and visualized the patterns using radar charts.

This research is unique in its focus on a video-based social media platform with a predominantly young user base. According to SimilarWeb (2024), the largest age group of visitors on Bilibili is 25–34-year-olds, while only 4% of users are over 65. Unlike YouTube, which has a more globally diverse audience and a broader range of content, Bilibili's content is heavily influenced by Chinese youth culture (Wang, 2022). This demographic distance from older adults provides an opportunity to investigate the younger generation's responses to the news framing of older adults.

**Literature review**

*Ageism in Media*

Originally defined by Butler (1980), ageism refers to stereotypes, prejudice, and discrimination against individuals or groups based on their age. It operates both explicitly and implicitly, affecting individuals, social networks, and broader institutional and cultural levels (Ayalon & Tesch-Römer, 2018). Ageism in media involves the discriminatory and stereotypical portrayal of older people, characterized by their underrepresentation, misrepresentation, and marginalization across traditional media, social media, and visual arts (Ivan et al., 2020; Loos & Ivan, 2018). Today, media plays a decisive role when it comes to shaping and staging images of aging (Wangler & Jansky, 2023).

Numeral studies have recorded the stereotypes and negative representation of elder adults in media, including a tendency to utilize certain stereotypes (Koskinen et al., 2014; Markov & Yoon, 2021), dense negative depictions (Ng, 2021; Lytle, Monahan, & Levy, 2023), and the focus on the vulnerabilities of the elderly, like helpless, frail, and unproductive, especially in emergent situations like health crises (Ayalon et al., 2021; Zhang & Liu, 2021; Reul et al., 2022; Sargu et al., 2023). The tendency of negative portrayal can inadvertently contribute to shaping societal views, thereby reinforcing, or even constructing biases toward the elderly (Ng, 2021; Edström, 2018; Xu, 2022).

Ageism in Chinese media, both social and traditional, is well-documented. For instance, Li (2021) found through qualitative textual analysis that older adults in China tend to portray themselves positively on social media, contrasting with the negative and stigmatizing images often presented by traditional media. Similarly, Zhang and Liu

(2021) revealed that media representation during the COVID-19 pandemic depicted older adults as vulnerable and passive recipients reliant on families, public institutions, and governments. This portrayal reinforced ageist stereotypes and intensified the dichotomized relationship between young and old, leading younger generations to perceive older adults as a 'threat' to public health.

**Framing as a theory of media effects**

Framing as a theory of media effects refers to the concept that the media draw attention to certain topics and place them within a field of meaning, which in turn influences audience perceptions and behaviors (Entman, 1993; Chong & Druckman, 2007; Aarøe, 2017). For framing effects of the older adults, the study by Wangler & Jansky (2023) found that presenting a negative age frame improved the self-image of old age but deteriorated the public image. Conversely, a positive frame improved the public image but decreased the self-image. It is opting to infer that the stereotypes that exist in media content could intense discrimination toward old people and reinforce self-stereotypes, further worsening the health or societal status of the old adults (Levy et al., 2000; Levy et al., 2008; Wheeler & Petty, 2001).

Previous research also underscores that certain frames on online platforms can lead to intensified emotional responses. These emotional reactions can, in turn, influence users' decisions to engage with content through comments, likes, or shares (Berger & Milkman, 2012; Oh, Bellur, & Sundar, 2015; Lavoie et al., 2021). The specific framing of content, therefore, plays a pivotal role in its reception and the extent to which it resonates with the platform's users. Media framing also has implications for financial interactions on platforms with reward or virtual reward systems. Users may be more inclined to financially back creators whose content either aligns with their pre-existing views or presents a novel or challenging perspective that they find valuable (Drivas et al., 2022; Scharlach & Hallinan, 2023). Understanding the nuances of framing on such platforms is paramount for determining if the popularity of specific media content related to older adults is linked to framing.

Based on the discussion of biased framing practice by media and online platforms, and its potential effects on public attitudes and behaviors toward older adults (Ng, 2021; Lytle, Monahan, & Levy, 2023; Ayalon et al., 2021; Zhang & Liu, 2021; Sargu et al., 2023; Reul et al., 2022), and depending on how content is framed can significantly influence the viewers' emotions, cognitions, and subsequent behaviors, such as sharing, commenting, or rewarding (Shahbaznezhad, Dolan, & Rashidirad, 2021; Tenenboim, 2022; Su, Liu, & McLeod, 2019; Cheonsoo & Yang, 2017), this study tries to build the link between different framings of news titles and engagement behavior metrics.

Previous research has shown that content evoking high-arousal emotions tends to be more viral (Berger & Milkman, 2012). Elements such as "negativity bias," "causal arguments," and "threats to personal or societal core values" can individually or collectively enhance the virality of social media messages (Mousavi et al., 2022). Therefore, we hypothesize that video news titles about older adults with negative framing will exhibit higher user engagement metrics.

**H1:** Video news titles with negative framing will gain more (1) views, (2) comments, (3) Danmaku, and (4) shares.

Considering that positive framing can evoke positive emotional responses and financial interactions (Drivas et al., 2022; Scharlach & Hallinan, 2023), we hypothesize the following:

**H2:** Video news titles with positive framing will gain more (5) likes, (6) rewards, and (7) collections.

**Frames Identified in Video News Titles Related to the Elderly**

We identified nine distinct frames by analyzing the original video news titles and reviewing related literature on the portrayal of older adults (Wangler & Jansky, 2023; Zhang & Liu, 2021; Ng & Indran, 2022; Amundsen, 2022). The two coders of this study coded all 2,017 video news titles according to the established coding book. Initially, a pilot coding was conducted on a subset of 200 titles. Using Cohen's alpha as a measure of inter-coder reliability, a value of α = .728 was achieved, indicating satisfactory agreement between the coders (O'Connor & Joffe, 2020). As shown in Table 1, these frames range from negative portrayals emphasizing vulnerability, marginalization, and criminal behavior (Frames 1-3), to positive portrayals highlighting achievements and enjoyment (Frames 4-5), and neutral stances focusing on policy, demographics, scientific insights, and factual reporting (Frames 6-9).

**Table 1**

*Details of video metrics*

| Variable | Min | Max | M | SD | N |
| --- | --- | --- | --- | --- | --- |
| Video Title | / | / | / | / | 2017 |
| Video Duration | 4 | 3039 | 95.73 | 142.28 | 1896 |
| (Uploader's) Fan Count | 0 | 7705500 | 118910.08 | 555160.65 | 2017 |
| View Count | 0 | 2550500 | 28887.50 | 117642.48 | 2017 |
| Like Count | 0 | 212500 | 1279.31 | 7876.77 | 2017 |
| Comment Count | 0 | 10129 | 187.86 | 666.99 | 2017 |
| Danmaku Count | 0 | 9490 | 49.13 | 323.78 | 2017 |
| Reward Count | 0 | 8919 | 56.54 | 390.19 | 2017 |
| Collection Count | 0 | 6673 | 68.64 | 322.387 | 2017 |
| Share Count | 0 | 6664 | 55.83 | 329.40 | 2017 |
| Frame Code | 1 | 9 | / | / | 2017 |

Negative portrayals include frames such as Physically or Mentally Frail, where older adults are depicted as weak or impaired, and Perpetrators of Crime, where they are shown as having breached legal boundaries. Positive portrayals are seen in frames like Accomplishing Ambitions, which highlight their achievements, and Embracing Enjoyment, showcasing their active engagement in enjoyable activities. Neutral frames include Policy Discourse, which examines policies impacting older adults, and Scientific Insights, offering research-based perspectives on aging.

**Methodology**

*Data Source*

This study utilizes a dataset from the news section of Bilibili, where only approved news channels in China can post video news. We chose news channels for their reliable, regulated content adhering to journalistic standards, allowing accurate analysis of media framing. The selection criteria stipulated that the titles must include the terms "老年人", "老人", "老年", or "长者". Furthermore, the video content had to be news reports addressing issues related to older adults, covering the period from 13 May 2016 to 25 January 2021. A total of 2,017 relevant video entries were identified based on these criteria.

*Video Metrics*

As shown in Table 2, for each video, we recorded news titles and various key metrics like views, likes, comments, and shares, as well as the duration of the video and the number of followers of the uploader. The analyzed video news metrics encompass:

*Video Title:* The headline of the video content.

*Video Duration:* The total playback time of the video, measured in seconds.

*Uploader's Fan Count:* The number of followers of the uploader, reflecting the uploader's popularity.

*View Count:* The total number of times the video has been viewed. It reflects a decision made before watching.

*Like Count:* The total number of "likes" a video has received.

*Comment Count:* The extent to which viewers engage in discussions about the video's content.

*Danmaku Count:* Danmaku is a commenting system that displays users' synchronous comments within the video stream, is widely used in Asian countries, especially in China and Japan (Lin, Huang, & Cordie, 2018; Teng & Chan, 2022).

The number of "bullet comments" or "danmaku", reflecting a form of interaction that is more active and immediate than comments.

*Reward Count:* The number of times viewers have rewarded or "tipped" the content creator for the video, representing a stronger appreciation and support for the content.

*Collection Count:* The number of viewers who have saved or "collected" the video for later viewing or reference, signifying that viewers find the content valuable and may wish to revisit it in the future.

*Share Count:* The number of times the video link has been shared outside the Bilibili platform, signaling its spread and potential influence beyond the immediate audience.

Table 2 provides descriptive statistics for video metrics across 2,017 videos. Key metrics include video duration (4-3,039 seconds, M=95.73), uploader's fan count (0-7.7 million, M=118,910.08), and view count (0-2.55 million, M=28,887.50). Engagement metrics such as like count (M=1,279.31), comment count (M=187.86), danmaku count (M=49.13), reward count (M=56.54), collection count (M=68.64), and share count (M=55.83) are also presented. Each video is categorized by a frame code ranging from 1 to 9.

**Table 2**

*The definition, example, and frequency of framings about older adults*

| Frames | Description | Example | Count |
|---|---|---|---|
| 1: Physically or Mentally Frail | depicting older adults as physically weak, vulnerable to illness, or mentally impaired | "How can an elderly person with unclear consciousness withdraw their savings?" 20/9/2018, 1818 Golden eyes, [老人神志不清，存款如何取出？20/9/2018，1818黄金眼] | 459 |
| 2: Background Figures | older adults are often relegated to ancillary roles within media narratives, serving merely to establish | "Salute! Beijing firefighters break the ice with their bodies to rescue a senior man in his sixties who fell into the water while winter fishing." 12/16/2020, Z | 553 |
| | context rather than being the central focus | news, [致敬!北京消防员用身体破冰，救援六旬冬钓落水老人，12/16/2020，中国蓝新闻] | |
| 3: Perpetrators of Crime | depict older adults as individuals who have breached legal or regulatory boundaries | "A 68-year-old individual set fire at Daming Lake Scenic Area out of curiosity...", 24/1/2021, New Express, [68 岁老人在大明湖景区放火：就是出于好奇…, 24/1/2021，新快报社] | 96 |
| 4: Accomplishing Ambitions | highlights older adults actively setting, pursuing, and achieving their previously unmet goals or aspirations, thereby countering ageist stereotypes | "Have you ever run a marathon? An 81-year-old completed a full marathon in 6 hours", 10/11/2020, China News Video. [你跑过马拉松吗？81 岁老人 6 小时跑完全马，10/11/2020，中新视频] | 156 |
| 5: Embracing Enjoyment | showcases seniors actively engaging in pleasurable activities, challenging the ageist perception that old age is synonymous with decline and discontent | "A 90-year-old celebrates their birthday while dancing joyfully to music and engaging in a dance-off with a young lady on stage from a distance", 9/10/2020, Ran News. [90 岁老人过生日，伴着音乐手舞足蹈，跟台上跳舞的小姐姐隔空斗舞，9/10/2020，燃新闻] | 215 |
| 6: Policy Discourse | examine policies and legislative measures that impact the lives of older adults, encompassing areas like social security, healthcare, and elder rights | "Green Light Extended for Elderly Pedestrians to Cross the Road - Hear What Those Present Have to Say", China Blue News, 28/9/2020. [绿灯为老人过马路延长，听听在场的他们怎么说，28/9/2020，中国蓝新闻] | 160 |
| 7: Demographic Trends | emphasizes the statistical and demographic patterns about the elderly population, providing valuable context that can enhance public awareness and understanding of aging societies | "In Hangzhou, one in every five people is elderly, and the number of female centenarians is twice that of males", Zhejiang Red TV, 2/8/2019. [杭州人5个里就有1个老人，百岁老人女的比男的多两倍！2/8/2019，浙样红 TV] | 4 |
| 8: Scientific Insights | the latest research on aging, offering a deeper comprehension of the biological, psychological, and social dimensions of the aging process | "Elderly Home Exercise｜Renowned Beijing Sport University Teacher Shows You Lower Limb Stretching", Xinhua Broadcasting, 16/3/2020. [老年人居家运动|北体大名师教您下肢拉伸，16/3/2020，新华广播] | 13 |
| 9: Factual Narration | emphasizes objective and detailed coverage of events, concentrating on the factual recounting of occurrences, essential details, and subsequent developments | "Breaking News! Thick smoke engulfs a residential area in Hangzhou city, an octogenarian tragically loses their life, 19/1/2020, Zheyang Hong TV". [突发！杭州市区一民居浓烟滚滚，八旬老人不幸遇难，19/1/ 2020，浙样红 TV] | 361 |

*Analytical Techniques*

To test the two hypotheses, an ANCOVA (Analysis of Covariance) was used to determine the influence of framing on user engagement metrics while accounting for potential confounding variables (Pallant, 2020). ANCOVA combines the principles of ANOVA and linear regression, enabling comparisons of one or more means while controlling for the variability of other factors (Egbewale, Lewis, & Sim, 2014). This method was chosen because several variables, such as the uploader's fan count and video duration, can significantly affect user engagement metrics. By using ANCOVA, we can isolate the impact of different frames on user engagement more accurately. The data were analyzed using IBM SPSS statistical software (IBM Corp., 2017). To illustrate the differences across nine distinct framings on individual user engagement metrics, radar figures were generated using Python 3.8 with the Matplotlib (version 3.3.4) and Seaborn (version 0.11.1) packages (Hunter, 2007; Waskom, 2021).

**Results**

*Correlation Between Framing and User Engagement Metrics*

Our first hypothesis proposes that video news titles with negative framing will result in higher (1) views, (2) comments, (3) Danmaku, and (4) shares. Our second hypothesis proposes that video news titles with positive framing will result in higher (5) likes, (6) rewards, and (7) collections. However, as shown in Table 3, the ANCOVA results reveal that framing significantly influences only views, shares, rewards, and collections.

Specifically, view count is significantly affected by the frame code, with an $F$ value of 3.40 and $p < .001$, suggesting that video framing can substantially impact viewership. The model explains 4.1% of the variance in view count ($R^2 = .041$). Share count is another engagement metric significantly influenced by the frame code, with an $F$ value of 3.07 and $p < .01$, indicating that the framing of a title could affect the likelihood of a video being shared. The model explains 6.1% of the variance in share count ($R^2 = .061$). Reward count and collection count also have strong correlations with the frame code, with $F$ values of 3.81 and 2.41, respectively, and $p$ values indicating statistical significance. The model explains 10.9% and 11.1% of the variance in reward and collection counts, respectively ($R^2$ = .109 and $R^2$ = .111). This implies that certain frames, potentially positive ones as proposed in the second hypothesis, may encourage viewers to reward and save the content more.

Regarding the two control factors shown in Table 3, the ANCOVA results also reveal that fan count is a significant predictor of nearly all user engagement metrics (views, likes, Danmakus, rewards, collections, and shares; all $p < .001$), highlighting the importance of an existing audience base in content interaction. Video duration significantly influences only the Danmaku count ($p < .05$), explaining 4.3% of the variance with fan counts ($R^2$ = .043).

Nevertheless, likes, comments, and Danmakus were not statistically significantly associated with the frame code. This suggests that although certain frames may affect viewers' decisions to view, share, reward, or collect content more substantially, they do not necessarily motivate users to like, comment on, or send Danmakus. To determine whether negative or positive framings truly amplify certain engagements, more detailed analysis and visualization are required.

**Table 3**

*ANCOVA results for multiple dependent variables*

| Dependent Variable | Predictor/Covariate | df | F | p | $R^2_{adj}$ |
|---|---|---|---|---|---|
|  | Fan Count *** | 1 | 61.29 | .000 |  |
| View Count | Video Duration | 1 | .81 | .367 | .041 |
|  | Frame Code *** | 8 | 3.40 | .000 |  |
| Like Count | Fan Count *** | 1 | 42.65 | .000 | .021 |
|  | Video Duration | 1 | .03 | .858 |  |
|  | Frame Code | 8 | 1.04 | .401 |  |
|  | Fan Count | 1 | .024 | .876 |  |
| Comment Count | Video Duration | 1 | 1.76 | .184 | .003 |
|  | Frame Code | 8 | 1.79 | .084 |  |
|  | Fan Count *** | 1 | 79.52 | .000 |  |
| Danmaku Count | Video Duration * | 1 | 4.12 | .042 | .043 |
|  | Frame Code | 8 | .91 | .504 |  |
|  | Fan Count *** | 1 | 200.5 | .000 |  |
| Reward Count | Video Duration | 1 | 3.92 | .531 | .109 |
|  | Frame Code *** | 8 | 3.81 | .000 |  |
|  | Fan Count *** | 1 | 218.59 | .000 |  |
| Collection Count | Video Duration | 1 | .055 | .814 | .111 |
|  | Frame Code * | 8 | 2.41 | .014 |  |
|  | Fan Count *** | 1 | 104.86 | .000 |  |
| Share Count | Video Duration | 1 | .534 | .465 | .061 |
|  | Frame Code ** | 8 | 3.07 | .002 |  |

Note. ***$p < .001$, **$p < .01$, *$p < .05$

*Different Engagement Patterns between Negative and Positive Framings*

错误!未找到引用源。 presents the selection of four viewer behaviors significantly related to framing code: views, shares, collections, and rewards, with their average values calculated across different frames.

**Views:** The average view count across all frames is 28,887.5. Frames depicting older adults as perpetrators of crime (Frame 3), providing factual narration (Frame 9), and depicting them as physically or mentally frail (Frame 1) have higher average view counts of 66,561.49, 41,766.19, and 37,583.61, respectively. These values are significantly above the overall mean, supporting our first hypothesis that negative framing results in higher views. Conversely, frames emphasizing scientific insights (Frame 8) and demographic trends (Frame 7) show lower view counts of 3,411.62 and 683, respectively, well below the mean, indicating less viewer engagement.

**Shares:** The average share count across all frames is 55.83. Frames 1 (physically or mentally frail), 3 (perpetrators of crime), and 9 (factual narration) exceed this average with share counts of 89.43, 71.58, and 106.87, respectively. This again supports the first hypothesis, suggesting that negative framing increases sharing behavior. On the other hand, Frame 2 (background figures) and Frame 6 (policy discourse) have lower share counts of 9.39 and 14.08, respectively, indicating less inclination to share these frames.

**Rewards:** The average reward count is 56.54. Frame 4 (accomplishing ambitions) and Frame 5 (embracing enjoyment) have higher reward counts of 222.12 and 88.66, respectively. This aligns with our second hypothesis that positive framing results in higher rewards. Frames with negative or neutral tones, such as Frame 6 (policy discourse) and Frame 7 (demographic trends), have lower reward counts of 21.78 and 2.25, respectively.

**Collections:** The average collection count is 68.64. Frame 4 (accomplishing ambitions) and Frame 5 (embracing enjoyment) again show higher collection counts of 127.64 and 96.04, respectively, supporting the second hypothesis. Frames like Frame 2 (background figures) and Frame 6 (policy discourse) exhibit lower collection counts of 29.45 and 24.41, respectively. However, Frame 3 (perpetrators of crime) also shows a higher collection count of 104.64, indicating that negative framing also has potential to save the content more. [错误!未找到引用源。 is here]

As shown in Figure 1, the radar chart further illustrates these findings, showing distinct patterns of user engagement across different frames. Negative frames like Frame 3 (perpetrators of crime) and Frame 1 (physically or mentally frail) prominently stand out with high views and share counts. In contrast, positive frames such as Frame 4 (accomplishing ambitions) and Frame 5 (embracing enjoyment) show higher counts in rewards and collections. Frame 9 (factual narration) is an exception, as it adheres to an objective reporting style while still generating high views and shares. Additionally, Frame 3, despite its negative connotation, also has the potential to boost collections. This visual representation reinforces the significant impact of framing on various user engagement metrics, highlighting the differing influences of negative and positive framing on viewer behavior.

**Figure 1**

*User engagement patterns across different framings*

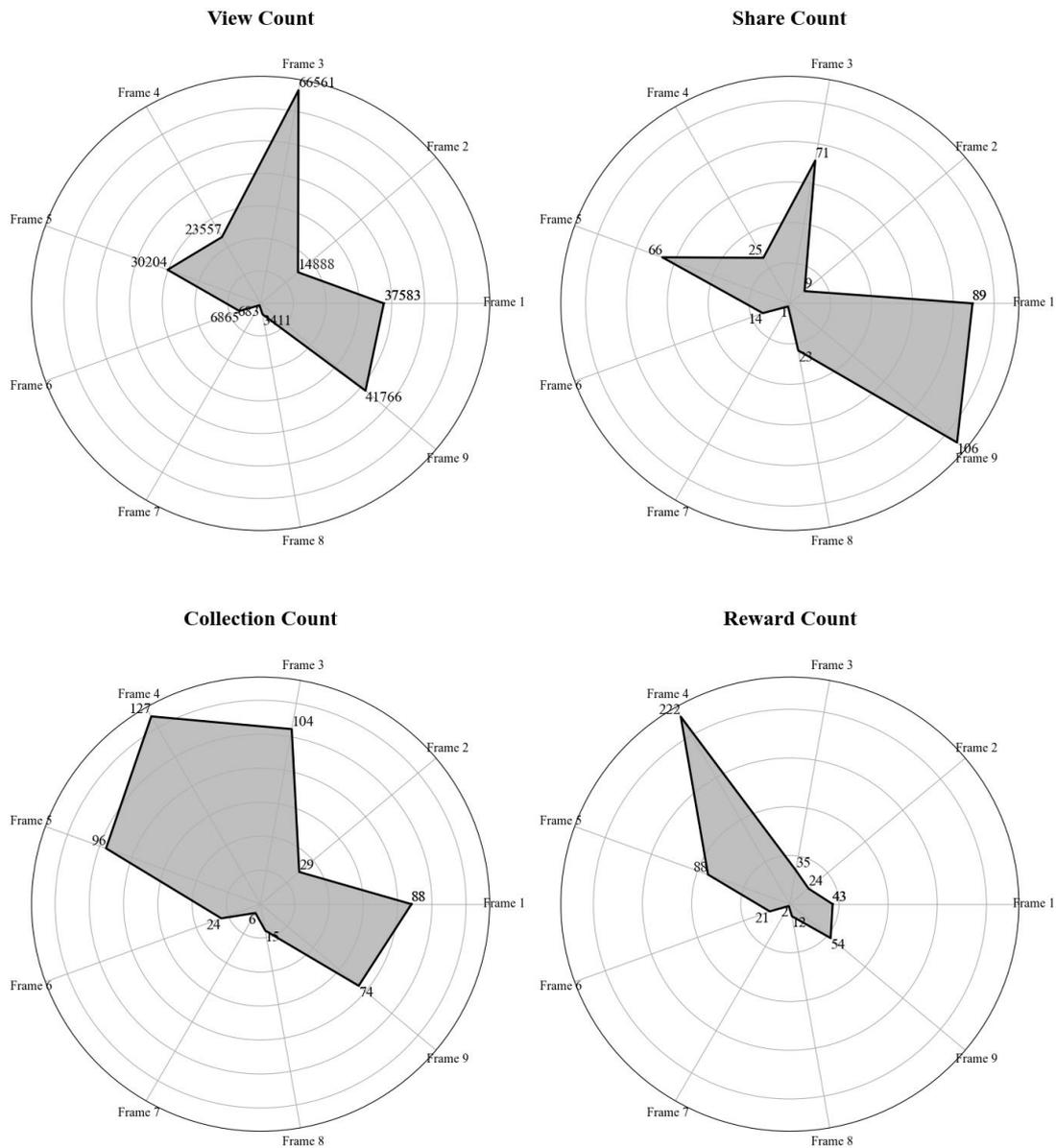

**Discussion**

The study reveals a link between how older adults are portrayed in video news titles and viewer behavior patterns. Different media frames resonate with audiences, influencing behaviors such as viewing, sharing, collecting, and rewarding content. Framing theory (Entman, 1993; Chong & Druckman, 2007; Aarøe, 2017; Borah, 2016) suggests that media presentation shapes audience interpretation and reaction. This research confirms that frames related to older adults impact user responses, highlighting framing's power in shaping video content engagement.

The observed pattern of high view and share counts in certain frames, such as Frames 1, 3, and 9, suggests a potential synergistic effect where high viewership may drive shares, and conversely, increased sharing could boost views. Viewing is the most fundamental user behavior, and the pronounced boost in views from negative framings (Frames 1 and 3) is notably significant. The corresponding high share rates for such videos warrant attention; these frames not only trigger viewing but, through sharing, can have broader implications across other platforms.

In contrast, positive frames (e.g., Frames 4 and 5) tend to inspire rewarding and collection behaviors, implying that these frames may instill a perception of value or importance, prompting users to reward or save the content. However, since these behaviors occur after viewing, the positive impact of these positive frames might be constrained without a substantial base of initial viewers.

Our findings contribute to the body of research indicating that media representations, including those concerning the older adults, influence social perceptions and behaviors (Wangler & Jansky, 2023; Makita et al., 2021). They enrich prior studies on the impact of framing on platform user behavior by investigating beyond mere popularity to explore the complex relationship between diverse user actions and the framing of video news titles about older adults (Berger & Milkman, 2012; Oh, Bellur, & Sundar, 2015; Lavoie et al., 2021). The study underscores the intricate and multifaceted nature of user engagement, shaped by the dynamic interplay between various frames and the users' perceptions and attitudes.

Importantly, the study's findings have significant implications for addressing ageism. The high engagement with negatively framed content about older adults suggests that such portrayals can perpetuate ageist stereotypes, reinforcing negative perceptions and potentially influencing how society views and treats older adults (Ng, 2021; Lytle, Monahan, & Levy, 2023; Ayalon et al., 2021; Zhang & Liu, 2021). Conversely, positive frames, while less effective in driving views and shares, encourage rewarding and collecting behaviors, indicating a perception of value and appreciation. This highlights the need for more balanced and positive media portrayals of older adults to combat ageism and promote a more respectful and inclusive societal view (World Health Organization, 2021).

As the global population ages, media platforms must be aware of the power they hold in shaping societal attitudes towards older adults. By promoting more positive and diverse representations, media can play a pivotal role in reducing ageism and fostering a more inclusive society (Langmann, 2023; Koskinen et al., 2014). Future research should continue to explore the impact of media framing on the perceptions of older adults, utilizing mixed methods and longitudinal studies to provide a deeper understanding of these dynamics.

**Limitations**

This study provides insights into the interplay between media framing related to older adults and user engagement but has limitations. The focus on Bilibili news video titles cannot fully apply to other platforms or content types where the portrayal of older adults and audience interactions differ. Content analysis's subjectivity could affect frame interpretation despite inter-coder reliability measures. User engagement metrics as proxies for societal perceptions may not fully capture audience attitudes and behaviors. Cross-sectional data limits causal inference, suggesting the need for longitudinal studies. Future research should broaden the scope to various media platforms and content types, use mixed methods to reduce subjectivity and employ longitudinal studies to better understand the evolving nature of audience engagement with media frames.

**Conclusion**

Through an analysis of 2,017 video news titles on Bilibili, this study demonstrates that negative frames, particularly those depicting older adults in a state of decline or as victims of sensational events, significantly increase views and shares. Conversely, positive portrayals of older adults encourage actions such as content collection and rewards, indicating a preference for more affirmative depictions. By applying framing theory, the research provides deeper insights into the dynamics of user engagement elicited by different representations of older adults. This study reinforces the significant influence of media framing on audience behavior and highlights the need for more equitable and dignified representations of older adults in media. Such representations are crucial for combating ageism and upholding the rights of older adults to a respectful portrayal in society. Future media practices should strive to balance portrayals to foster a more inclusive and respectful view of the aging population.